\documentclass{article}
\usepackage{arxiv}

\usepackage[utf8]{inputenc}
\usepackage[T1]{fontenc}
\usepackage{url}
\usepackage{booktabs}
\usepackage{amsmath,amssymb,amsfonts,mathtools}
\usepackage{amsthm}
\usepackage{nicefrac}
\usepackage{microtype}
\usepackage{graphicx}
\usepackage{xcolor}
\usepackage{subcaption}
\usepackage[shortlabels]{enumitem}
\usepackage{pifont}
\usepackage{tcolorbox}
\tcbuselibrary{skins,breakable}
\usepackage{cite}
\usepackage{hyperref}
\hypersetup{colorlinks=true,linkcolor=blue,citecolor=blue,urlcolor=blue}

\tcbset{
  problembox/.style={
    enhanced, breakable,
    colback=blue!3!white, colframe=blue!40!black,
    boxrule=0.6pt, arc=1.5mm,
    left=2mm, right=2mm, top=1.5mm, bottom=1.5mm
  }
}

\newcommand{\R}{\mathbb{R}}
\newcommand{\N}{\mathbb{N}}
\newcommand{\SM}{S^{\chi}_{M}}

\newtheorem{theorem}{Theorem}[section]
\newtheorem{assumption}[theorem]{Assumption}
\newtheorem{proposition}[theorem]{Proposition}
\newtheorem{Definition}[theorem]{Definition}
\newtheorem{lemma}[theorem]{Lemma}

\newtheorem{problem}[theorem]{Problem}

\title{Incremental Input-to-State Stabilization of Partially Unknown
Control-Affine Systems within Prescribed Safe Sets}

\author{
  P Sangeerth \\
  Centre for Cyber-Physical Systems\\
  Indian Institute of Science\\
  Bengaluru, India\\
  \texttt{sangeerthp@iisc.ac.in}
  \And
  Poulomee Ghosh \\
  Centre for Cyber-Physical Systems\\
  Indian Institute of Science\\
  Bengaluru, India\\
  \texttt{poulomeeg@iisc.ac.in}
  \And
  Pushpak Jagtap \\
  Centre for Cyber-Physical Systems and\\
  Department of Aerospace Engineering\\
  Indian Institute of Science, Bengaluru, India\\
  \texttt{pushpak@iisc.ac.in}
}
 
\begin{document}

\maketitle
\pagestyle{empty}

\begin{abstract}
Incremental input-to-state stability ($\delta$-ISS) guarantees that the trajectories of a system converge towards one another, up to a gain on the difference of their inputs. Some existing learning-based designs that combine incremental stability with safety achieve only incremental input-to-state practical stability, whose non-vanishing
offset can make the guarantee conservative. We consider control-affine systems with a known, full-row-rank input matrix and an unknown drift learned using Gaussian process regression. The prescribed safe set is a superlevel set of a smooth function satisfying a radial condition that ensures a diffeomorphism onto $\mathbb R^n$ and it need not contain
any equilibrium of the uncontrolled system. Using this transformation, we design a smooth feedback law and derive sufficient conditions that, under the stated learning assumptions, guarantee with high probability both forward invariance of a compact subset of the safe set and $\delta$-ISS on that subset. The incremental estimate uses the distance induced by the transformation and contains no practical offset. Simulations on a Van der Pol oscillator and a Lotka--Volterra
system illustrate the proposed approach.
\end{abstract}

\section{Introduction}

Incremental stability is a property of nonlinear systems characterized
by the convergence of system trajectories with respect to one another,
rather than to a fixed equilibrium point or nominal trajectory. Various
analytical frameworks have been developed to study this property, with
contraction analysis~\cite{Lohmiller15062005, sun2021learning,
TSUKAMOTO2021135} and convergent dynamics~\cite{pavlov2006uniform}
serving as prominent tools. Incremental Lyapunov functions have emerged
as a versatile mechanism for analysing incremental stability~\cite{angeli2002lyapunov,
backsteppingzamani}, and have been extended to general nonlinear
systems~\cite{angeli2002lyapunov}, stochastic and hybrid
dynamics~\cite{8350276}, time-delay systems~\cite{chaillet2013razumikhin},
and interconnected switched networks~\cite{dey2026incremental} including Math2Real controller transfer \cite{sangeerth2026transferring}. Among
these notions, incremental input-to-state stability ($\delta$-ISS) is
particularly important, as it guarantees robust convergence in the
presence of external inputs~\cite{zamani2017towards}, and state-feedback
controllers that render known nonlinear systems $\delta$-ISS have been
proposed in~\cite{pushpakHamilton, backsteppingzamani}.

For systems with unknown dynamics, establishing $\delta$-ISS requires
accounting for model uncertainty while controlling the distance between
trajectories. This becomes more challenging when each trajectory must
also remain within a prescribed safe set. Learning-based $\delta$-ISS designs have been
proposed for unknown systems, but they rely on restrictive structural
assumptions or yield conservative guarantees. The approach
in~\cite{11312975} synthesizes a $\delta$-ISS controller from trajectory
data, but is limited to polynomial dynamics without ensuring safety, while the neural-network
approach in~\cite{basu2026neural} achieves $\delta$-ISS within a compact
set at the cost of significant conservative bounds and training resources. In~\cite{sangeerth2025controller}, the
unknown dynamics of Euler-Lagrange systems are learned using Gaussian
processes (GPs), and invariance is enforced using control barrier
functions (CBFs). However, this design only guarantees incremental
input-to-state \emph{practical} stability ($\delta$-ISpS): the model
error and the CBF correction terms are absorbed into a non-vanishing
offset, which can be large enough to render the guarantee of limited
practical use.

The difficulty in establishing invariance and $\delta$-ISS together is that invariance concerns each trajectory individually, whereas $\delta$-ISS requires a bound on the distance between any two trajectories. In a CBF-based design, the control objective is to ensure safety for individual trajectories. The incremental analysis must therefore account for the difference between the two control actions. Bounding this difference by a constant can introduce a non-vanishing offset in the incremental estimate. Approaches that merge Lyapunov and barrier conditions into a single
certificate, such as Lyapunov-barrier functions~\cite{meng2025zubov},
certify convergence to an equilibrium and therefore assume that this
equilibrium lies inside the safe set. 

In this paper, we address both the
limitations by embedding the safety requirement into the coordinates in
which incremental stability is analysed, without requiring the safe set
to contain an equilibrium. We construct a diffeomorphism that maps the
safe set onto $\R^n$ and blows up at its boundary, so that boundedness in
the new coordinates is equivalent to invariance in the original ones,
while incremental stability is established with respect to the distance
induced by the same map. In this work, we consider systems whose input matrix has full row rank. Even so, the problem does not reduce to cancelling the drift, because
the learned model is reliable only on the region where data are collected, and the residual model error is amplified by the derivative of the diffeomorphism, which grows unbounded near the boundary of the safe set. The controller must therefore dominate this amplified error so that both trajectories remain in a compact subset of the safe set, while still bounding the distance between them.

The main contributions of this paper are as follows.
\begin{enumerate}[1)]
    \item For control-affine systems with a known input matrix and an
    unknown drift learned via GP regression, we design a smooth feedback
    law that keeps the state in a prescribed safe set by rendering a compact subset of it forward invariant. The safe set
    is a superlevel set of a user-defined function and is only required to
    be star-shaped with respect to an interior point; in particular, it
    need not contain any equilibrium of the open-loop dynamics.
    \item We establish, with high probability, $\delta$-ISS of the closed-loop system with respect to the distance induced by the transformation. The estimate depends only on the initial distance and the input mismatch. Thus, for any two admissible trajectories driven by the same input, the controller guarantees convergence towards one another while both remain in the safe set.
    
\end{enumerate}

\textbf{Notations: }
We denote the real, positive real, and non-negative real numbers by
$\R$, $\R^+$, and $\R^+_0$, and the positive integers by $\N$.
The norm $\|\cdot\|$ is the Euclidean vector norm or induced matrix
two-norm; $A^\top$ and $A^\dagger$ denote the transpose and
Moore--Penrose pseudoinverse of $A$, and $\mathbb{I}$ is the identity
matrix of appropriate dimension. For measurable
$\upsilon:\R^+_0\to\R^n$,
$\|\upsilon\|_\infty:=\operatorname{ess\,sup}_{t\ge0}\|\upsilon(t)\|$.
The closure and boundary of a set $A$ are $\bar A$ and $\partial A$.
A continuous $\alpha:\R^+_0\to\R^+_0$ belongs to $\mathcal{K}$ if it
is strictly increasing and $\alpha(0)=0$, and to $\mathcal{K}_\infty$
if also unbounded. A continuous
$\beta:\R^+_0\times\R^+_0\to\R^+_0$ belongs to $\mathcal{KL}$ if
$\beta(\cdot,t)\in\mathcal{K}$ for each $t$ and
$\beta(r,t)\to0$ monotonically as $t\to\infty$ for each $r$.
A metric $\mathbf{d}:\R^n\times\R^n\to\R^+_0$ is symmetric,
satisfies the triangle inequality, and obeys
$\mathbf{d}(x,y)=0$ if and only if $x=y$.

\section{Incremental Input-to-State Stability}\label{sec: delta_iss}

Consider the system $\dot x=\mathcal F(x,u)$, with state
$x\in\R^n$, control input $u\in\R^m$, and locally Lipschitz
vector field $\mathcal F$. Under the feedback law $u=p(x,\upsilon)$,
the closed-loop dynamics are
\begin{equation}\label{eqn:sys_cl}
    \dot x=f_p(x,\upsilon)
    :=\mathcal F\bigl(x,p(x,\upsilon)\bigr),
\end{equation}
where the external input $\upsilon$ takes values in
$W:=\{w\in\R^n:\|w\|\le\bar\upsilon\}$ for some user-defined
$\bar\upsilon\ge0$. We assume that $f_p$ is continuous and
locally Lipschitz in $x$, uniformly over $W$, on its domain.
An admissible input is a measurable signal
$\upsilon:\R_0^+\to W$. We denote by $x_{a\upsilon}$ the
solution of~\eqref{eqn:sys_cl} from $x(0)=a$ under $\upsilon$.

\begin{Definition}[Forward invariance]\label{def:invariance}
Set $A$ is forward invariant under~\eqref{eqn:sys_cl} if, for
every $a\in A$ and every admissible input $\upsilon$, the solution
$x_{a\upsilon}$ exists for all $t\ge0$ and remains in $A$.
\end{Definition}

We use the following metric formulation of incremental
input-to-state stability~\cite{angeli2002lyapunov,backsteppingzamani}.

\begin{Definition}[$\delta$-ISS on a set $A$]\label{def:deltaISS}
Let $A$ be forward invariant under~\eqref{eqn:sys_cl}, and let
$\mathbf d$ be a metric on $A$. The system is incrementally
input-to-state stable ($\delta$-ISS) on $A$ with respect to
$\mathbf d$ if there exist $\beta\in\mathcal{KL}$ and
$\gamma\in\mathcal K_\infty$ such that, for all $a,a'\in A$,
all admissible inputs $\upsilon,\upsilon'$, and all $t\ge0$,
\begin{align}\label{eqn:deltaISS}
    \mathbf d\bigl(x_{a\upsilon}(t),x_{a'\upsilon'}(t)\bigr)
    \le \beta\bigl(\mathbf d(a,a'),t\bigr)
       +\gamma\bigl(\|\upsilon-\upsilon'\|_\infty\bigr),
\end{align}
where $\|\cdot\|_\infty$ denotes the essential supremum norm.
\end{Definition}

If~\eqref{eqn:deltaISS} holds with an additional constant $c>0$
on the right-hand side, the system is incrementally
input-to-state practically stable
($\delta$-ISpS)~\cite{sangeerth2025controller}. For identical
inputs, the $\delta$-ISS estimate decays to zero, whereas
practical estimate retains the offset $c$.

\begin{Definition}[$\delta$-ISS Lyapunov function on $A$]
\label{def:deltaISSLyapunov}
Let $A$ be forward invariant under~\eqref{eqn:sys_cl}, and let
$\mathbf d$ be a metric on $A$. A continuously differentiable function
$V:A\times A\to\R_0^+$, is a $\delta$-ISS Lyapunov function with respect to
$\mathbf d$ if there exist
$\underline\alpha,\overline\alpha,\sigma\in\mathcal K_\infty$
and $\kappa>0$ such that, for all $x,x'\in A$ and $w,w'\in W$,
\begin{enumerate}[(i)]
    \item\label{item:lyapunov1}
    $\underline\alpha\bigl(\mathbf d(x,x')\bigr)
    \le V(x,x')
    \le \overline\alpha\bigl(\mathbf d(x,x')\bigr)$,
    \item\label{item:lyapunov2}
    $D_1V(x,x')f_p(x,w)+D_2V(x,x')f_p(x',w')
    \le -\kappa V(x,x')+\sigma(\|w-w'\|)$,
\end{enumerate}
where $D_1V$ and $D_2V$ denote the derivatives with respect to
the first and second arguments.
\end{Definition}

The following lemma gives a sufficient condition for $\delta$-ISS on a forward invariant set.

\begin{lemma}[\cite{basu2026neural}]\label{thm:lyapunov_char}
Let $A$ be forward invariant under~\eqref{eqn:sys_cl}, and let
$\mathbf d$ be a metric on $A$. If the closed-loop system admits
a $\delta$-ISS Lyapunov function on $A$ with respect to $\mathbf d$,
then the system is $\delta$-ISS on $A$ with respect to $\mathbf d$.
\end{lemma}

\section{Problem Formulation}\label{sec:problem}


\subsection{System and Data}

Consider the control-affine system
\begin{equation}\label{eqn:sys_affine}
    \dot x=f(x)+g(x)u,
\end{equation}
where $x\in\R^n$ is the state, $u\in\R^m$ is the control input,
and $m\ge n$.

\begin{assumption}\label{ass:system}
Let $\mathcal D\subseteq\mathbb R^n$ be an open set containing
$\overline S$. The drift $f:\mathcal D\to\mathbb R^n$ is unknown
and of class $C^2$. The input matrix
$g:\mathcal D\to\mathbb R^{n\times m}$ is known, of class $C^2$,
and has full row rank for every $x\in\mathcal D$.
\end{assumption}

The rank condition gives $g(x)g(x)^\dagger=I_n$, where
$g(x)^\dagger$ denotes the Moore--Penrose pseudoinverse.
The feedback law has the form $u=p(x,\upsilon)$, where
$\upsilon$ is the admissible external input defined in
Section~\ref{sec: delta_iss}.

\subsection{Safe Set}\label{subsec:safe_set}

The state is required to remain in a prescribed safe set defined by
a $C^2$ function $h:\R^n\to\R$ as
\begin{equation}\label{eqn:safe_set}
    S:=\{x\in\R^n\mid h(x)>0\}.
\end{equation}
For an interior point $x_c\in S$, consider the transformation
\begin{equation}\label{eqn:chi}
    \chi(x):=\frac{x-x_c}{h(x)}, \qquad x\in S.
\end{equation}
As $x$ approaches any $x_b\in\partial S$ from within $S$,
continuity of $h$ gives $h(x)\to 0$, while
$\|x-x_c\|\to\|x_b-x_c\|>0$. Hence $\|\chi(x)\|\to\infty$.
The transformation also provides a way to compare trajectories
through the distance
\[
    \mathbf{d}(x,x'):=\|\chi(x)-\chi(x')\|, \qquad x,x'\in S.
\]
For $\mathbf{d}$ to define a metric on $S$, distinct states must
have distinct images under $\chi$.

To see why the geometry of $S$ matters, consider two examples in
$\R^2$. If $S$ is a disc and $x_c$ is any interior point, each
ray from $x_c$ meets $\partial S$ exactly once. If $S$ is an
annulus, a ray from $x_c$ through the hole leaves $S$ and
re-enters it on the opposite side. Points on these two portions
of the ray can have the same image under $\chi$. The geometric
property that distinguishes the two examples is
\emph{star-shapedness}: $S$ is star-shaped with respect to $x_c$
if the line segment from $x_c$ to every $x\in S$ lies in $S$.
Even on a star-shaped set, injectivity of $\chi$ depends on how
$h$ varies along each ray.

To obtain a sufficient condition, write $x=x_c+t\nu$, where
$t>0$ and $\|\nu\|=1$, and let $G(t):=h(x_c+t\nu)/t$.
Along the portion of the ray in $S$, we have
$\chi(x_c+t\nu)=\nu/G(t)$. Thus, if $G$ is strictly decreasing,
the transformed radius $1/G(t)$ is strictly increasing.
Differentiation gives $G'(t)=-\phi(x)/t^2$, where
\begin{equation}\label{eqn:phi}
    \phi(x):=h(x)-\nabla h(x)^\top(x-x_c).
\end{equation}

\begin{assumption}\label{ass:safe_set}
The set $S$ is bounded, and there exists $x_c\in S$ such that
$\phi(x)>0$ for all $x\in S$.
\end{assumption}

Under Assumption~\ref{ass:safe_set}, $S$ is star-shaped with respect
to $x_c$, and $\chi$ is a bijection from $S$ onto $\R^n$ with invertible
Jacobian; hence $\mathbf{d}$ is a metric on $S$ (Lemma~\ref{lem:chi}).

For a bounded safe set, concavity of $h$ provides a simple
sufficient condition for Assumption~\ref{ass:safe_set}. Indeed,
for any $x_c\in S$, the tangent inequality gives
\[
    \phi(x)
    =h(x)+\nabla h(x)^\top(x_c-x)
    \ge h(x_c)>0, \qquad x\in S.
\]
This includes ellipsoids and superellipsoids with concave $C^2$
defining functions. The point $x_c$ serves only as the center of
the transformation. In particular, $S$ need not contain an
equilibrium of the open-loop drift $f$.

\subsection{Problem Statement}

For a prescribed $M>0$, define
\begin{equation}\label{eqn:SM}
    \SM:=\{x\in S:\|\chi(x)\|\le M\}.
\end{equation}
Under Assumption~\ref{ass:safe_set}, $\SM$ is the image of the
closed ball of radius $M$ under $\chi^{-1}$ and is therefore a
compact subset of $S$. These sets are nested, and their union
over $M>0$ is $S$. Thus $M$ specifies the region of initial
states for which safety and incremental stability are required.

\begin{tcolorbox}[problembox]
\begin{problem}\label{prob:main}
Consider system~\eqref{eqn:sys_affine} and the safe
set~\eqref{eqn:safe_set} under
Assumptions~\ref{ass:system} and~\ref{ass:safe_set}.
Given $M>0$,
$\epsilon\in(0,1)$, and noisy measurements of the drift,
design a feedback law $u=p(x,\upsilon)$ such that, with
probability at least $(1-\epsilon)^n$, both of the following hold:
\begin{enumerate}[(i)]
    \item $\SM$ is forward invariant for every admissible
    external input $\upsilon$.
    \item The closed-loop system is $\delta$-ISS on $\SM$
    with respect to the metric $\mathbf d$.
\end{enumerate}
\end{problem}
\end{tcolorbox}
\section{Gaussian Process Model}\label{sec:gp}

To deal with unknown parts of the dynamics, we use a GP approximation, which requires the data. To learn the drift, we use
measurements $\{(x^{(j)},y^{(j)})\}_{j=1}^N$ of the form
\[
    y^{(j)}=f(x^{(j)})+\omega^{(j)},
\]
where $\omega^{(j)}$ denotes measurement noise. The assumptions
on the data and the GP model are given as follows. 

\begin{assumption}[\cite{HircheGP}]\label{assume:RKHS}
There exists an open convex set $\hat S\supset\bar S$ such that each
component $f_i$, restricted to $\hat S$, belongs to the RKHS of a known
kernel $k_i$, with $\lVert f_i\rVert_{k_i}\le B_i$ for a known constant
$B_i$.
\end{assumption}

Since $S$ is bounded, a set $\hat{S}$ always exists. For instance, $\hat S$
can be taken as any open ball containing $\bar S$. The set $\hat S$ is
used only in the analysis: since $S$ is star-shaped but not necessarily
convex, a convex set containing $\bar S$ is needed to bound the
variation of the model error in Lemma~\ref{lem:lipschitz}. To train the model, we also need the following assumption on
the availability of data.

\begin{assumption}[\cite{HircheGP}]\label{assume:measurements}
Measurements $x\in\bar S$ and $y=f(x)+\omega$ are available, where
$\omega\sim\mathcal{N}(0_n,\rho_f^2\mathbb{I}_n)$ is additive noise, independent across measurements and of the measurement locations.
\end{assumption}

\subsection{Gaussian Process Models}\label{section:GP}

Gaussian processes (GPs) can be used to learn an unknown nonlinear
function from noisy measurements, while also quantifying the model
uncertainty based on the distance to the training data~\cite{GPBook}.
Since GP regression is scalar-valued, each component $f_i$ of $f$,
$i\in\{1,\ldots,n\}$, is modelled by a separate GP with zero prior mean
and squared-exponential kernel
\begin{align}\label{eqn:kernel}
    k_i(x,x')=s_i^2\exp\Big(-\sum_{j=1}^n\frac{(x_j-x'_j)^2}{2l_{ij}^2}\Big),
    \qquad x,x'\in\hat S,
\end{align}
where $s_i>0$ is the signal amplitude and $l_{ij}>0$ is the length scale
of $k_i$ along the $j$-th state coordinate. These hyperparameters are
typically obtained by maximizing the marginal likelihood~\cite{GPBook}, and are then kept fixed when computing the bounds below.

Let $\mathcal{D}=\{(x^{(j)},y^{(j)})\}_{j=1}^N$ be a data set satisfying
Assumption~\ref{assume:measurements}, i.e., $x^{(j)}\in\bar S$ and
$y^{(j)}=f(x^{(j)})+\omega^{(j)}$. For each $i$, collect
$y_i:=[y_i^{(1)},\ldots,y_i^{(N)}]^\top\in\R^N$, the Gram matrix
$K_i\in\R^{N\times N}$ with entries $[K_i]_{jl}=k_i(x^{(j)},x^{(l)})$, and
$\bar k_i(x):=[k_i(x,x^{(1)}),\ldots,k_i(x,x^{(N)})]^\top\in\R^N$.
The posterior mean and variance of $f_i$ at $x\in\hat S$ are
\begin{align}
    \mu_i(x)&=\bar k_i(x)^\top\theta_i,\qquad
    \theta_i:=(K_i+\rho_f^2\mathbb{I}_N)^{-1}y_i,\label{eqn:mean_i}\\
    \rho_i^2(x)&=k_i(x,x)-\bar k_i(x)^\top(K_i+\rho_f^2\mathbb{I}_N)^{-1}\bar k_i(x),\label{eqn:variance_i}
\end{align}
where $\theta_i\in\R^N$ is the weight vector of the posterior mean. We
write $\mu(x):=[\mu_1(x),\ldots,\mu_n(x)]^\top$ for the learned drift and
$e(x):=f(x)-\mu(x)$ for the model error. Since $\rho_i$ is continuous
and $\bar S$ is compact, $\bar\rho_i:=\max_{x\in\bar S}\rho_i(x)$ exists. The magnitude
of the model error can then be bounded with high probability as
follows.


\begin{proposition}\label{prop:gp}
Let Assumptions~\ref{assume:RKHS} and~\ref{assume:measurements} hold,
let $\epsilon\in(0,1)$, and $\eta_i:=B_i+\sqrt{\ln\det\big(\mathbb{I}_N+\rho_f^{-2}K_i\big)+2\ln(1/\epsilon)}$.
Then there is an event $\mathcal{E}:=\bigcap_{i=1}^n\mathcal{E}_i$ with $\mathcal{E}_i=\left\{| f_i(x)-\mu_i(x)| \leq |\eta_i| |\rho_i(x)|,\forall x\in X\right\}$, $\mathbb{P}(\mathcal{E})\ge(1-\epsilon)^n$ for all $i$,
and hence, with $\bar e:=\big(\sum_{i=1}^n\eta_i^2\bar\rho_i^2\big)^{1/2}$,
\begin{align}\label{eqn:gp_bound}
    \mathbb{P}\big\{\lVert e(x)\rVert\le\bar e,\ \forall x\in\bar S\big\}\ge(1-\epsilon)^n.
\end{align}
\end{proposition}

\begin{proof}
We refer to~\cite[Theorem~3.11]{abbasi2012online}, applied to each
component $f_i$ with regularization $\rho_f^2$, which gives an event $\mathcal{E}_i$ with $\mathbb{P}(\mathcal{E}_i)\ge1-\epsilon$ on which $|e_i(x)|\le\eta_i\rho_i(x)$ for all $x\in\hat S$. Since the events $\mathcal{E}_i$ depend on different, independent noise components, $\mathcal{E}:=\bigcap_{i=1}^n\mathcal{E}_i$ satisfies $\mathbb{P}(\mathcal{E})\ge(1-\epsilon)^n$, and on $\mathcal{E}$, $\lVert e(x)\rVert^2\le\sum_i\eta_i^2\rho_i^2(x)\le\bar e^2$ for all $x\in\bar S$.
\end{proof}

\subsection{Lipschitz Bound on the Model Error}

Proposition~\ref{prop:gp} bounds magnitude of the model error. Our
analysis also requires a bound on how fast $e$ can vary. To derive it, we use the following
mean value inequality.

\begin{lemma}[Mean value inequality]\label{lem:mvt}
Let $C\subseteq\R^p$ be convex and let $\Phi$ be a continuously
differentiable vector- or matrix-valued map on an open set containing
$C$. Then, for all $a,b\in C$,
$\lVert\Phi(a)-\Phi(b)\rVert\le\big(\sup_{z\in C}\lVert\tfrac{\partial\Phi}{\partial z}(z)\rVert\big)\lVert a-b\rVert$.
\end{lemma}

\begin{proof}
The segment $\zeta(s):=b+s(a-b)$, $s\in[0,1]$, lies in $C$ by convexity.
By the fundamental theorem of calculus and the chain rule,
$\Phi(a)-\Phi(b)=\int_0^1\frac{\partial\Phi}{\partial z}(\zeta(s))(a-b)\,ds$.
Taking norms and bounding the integrand by its supremum over $C$ gives
the result.
\end{proof}

We also need to know the Lipschitz constant of the model error $e$ on $\bar S$, for which we use the following results. We know that the derivative of a GP with a $C^2$ kernel is again a GP. For the
kernel~\eqref{eqn:kernel}, the posterior variance of
$\partial f_i/\partial x_j$ at $x\in\hat S$ is
\begin{align}\label{eqn:deriv_var}
    \rho_{\partial,ij}^2(x)=\frac{s_i^2}{l_{ij}^2}
    -\partial_j\bar k_i(x)^\top(K_i+\rho_f^2\mathbb{I}_N)^{-1}\partial_j\bar k_i(x),
\end{align}
where $\partial_j\bar k_i(x)\in\R^N$ collects the entries
$\frac{\partial k_i(x,x^{(l)})}{\partial x_j}=-\frac{x_j-x^{(l)}_j}{l_{ij}^2}k_i(x,x^{(l)})$.
The derivative of the model error is then bounded on the same event as
in Proposition~\ref{prop:gp}.

\begin{proposition}\label{prop:gp_deriv}
Let Assumptions~\ref{assume:RKHS} and~\ref{assume:measurements} hold,
and let $\eta_i$ and $\mathcal{E}$ be as in Proposition~\ref{prop:gp}.
Then, on $\mathcal{E}$,
\begin{align}\label{eqn:gp_deriv_bound}
    \Big|\frac{\partial e_i}{\partial x_j}(x)\Big|\le\eta_i\,\rho_{\partial,ij}(x)
\end{align}
for all $x\in\hat S$ and all $i,j\in\{1,\ldots,n\}$.
\end{proposition}

\begin{proof}
See Appendix~\ref{appendix: proof_prop_gp_deriv}.
\end{proof}
Using Proposition~\ref{prop:gp_deriv}, the Lipschitz constant of $e$ is
obtained in the following lemma.
\begin{lemma}\label{lem:lipschitz}
Let $\bar\rho_{\partial,ij}:=\max_{x\in\mathrm{conv}(\bar S)}\rho_{\partial,ij}(x)$,
where $\mathrm{conv}(\bar S)\subset\hat S$ is the convex hull of $\bar S$.
On the event $\mathcal{E}$ of Proposition~\ref{prop:gp},
$\lVert e(x)-e(x')\rVert\le L_e\lVert x-x'\rVert$ for all
$x,x'\in\mathrm{conv}(\bar S)$, and in particular for all
$x,x'\in\bar S$, where
\begin{align}\label{eqn:Le}
    L_e:=\Big(\sum_{i=1}^n\sum_{j=1}^n\eta_i^2\,\bar\rho_{\partial,ij}^2\Big)^{1/2}.
\end{align}
\end{lemma}

\begin{proof}
By~\eqref{eqn:gp_deriv_bound}, the Frobenius norm, and hence the
induced norm, of $\frac{\partial e}{\partial x}(x)$ is at most $L_e$ on
$\mathrm{conv}(\bar S)$, and the result follows from Lemma~\ref{lem:mvt}
with $C=\mathrm{conv}(\bar S)$.
\end{proof}
\section{Main Result}\label{sec:main}

We now design a feedback law, based on the learned drift $\mu$, that
renders $\SM$ forward invariant and the closed-loop system $\delta$-ISS
on $\SM$. The analysis is carried out in the coordinates
$z:=\chi(x)\in\R^n$, in which $\partial S$ is mapped to infinity. We write
$w:=x-x_c$ and $J(x):=\partial\chi(x)/\partial x$. The following lemma
collects the properties of $\chi$ needed to pass between $x\in S$ and
$z\in\R^n$.

\begin{lemma}\label{lem:chi}
Under Assumption~\ref{ass:safe_set}, the following hold:
\begin{enumerate}[(i)]
    \item for every $x\in S$, $J(x)$ is invertible, with
    \begin{align}
        J(x)&=\frac{1}{h(x)}\Big(\mathbb{I}-\frac{w\,\nabla h(x)^\top}{h(x)}\Big),\label{eqn:J}\\
        J(x)^{-1}&=h(x)\Big(\mathbb{I}+\frac{w\,\nabla h(x)^\top}{\phi(x)}\Big);\label{eqn:Jinv}
    \end{align}\label{item: lemma_chi_i}
    \item $\chi$ is a bijection from $S$ onto $\R^n$;\label{item: lemma_chi_ii}
    \item the inverse $\psi:=\chi^{-1}:\R^n\to S$ is $C^1$, with
    $\frac{\partial\psi}{\partial z}(z)=J(\psi(z))^{-1}$.\label{item: lemma_chi_iii}
\end{enumerate}
\end{lemma}

\begin{proof}
See Appendix~\ref{appendix: proof_lem_chi}.
\end{proof}

Thus $\mathbf{d}(x,x')$ is the Euclidean distance between $z=\chi(x)$
and $z'=\chi(x')$, and $\SM=\psi(B_M)$, where
$B_M:=\{z\in\R^n\mid\lVert z\rVert\le M\}$.


Now we present the main result of the paper, i.e., a feedback
law that renders the set $\SM$ forward invariant and the closed-loop
system $\delta$-ISS on $\SM$.

\begin{theorem}\label{thm:main}
Consider system~\eqref{eqn:sys_affine} under
Assumptions~\ref{ass:system}, \ref{ass:safe_set}, \ref{assume:RKHS},
and~\ref{assume:measurements}, with $f$ approximated by a GP with mean
$\mu$ given by~\eqref{eqn:mean_i}. For given $\epsilon\in(0,1)$ and
$M>0$, the feedback law
\begin{align}\label{eqn:ctrl}
u\hspace{-0.25em}=\hspace{-0.25em}g(x)^\dagger\Big[&J(x)^{-1}\hspace{-0.25em}\big(\hspace{-0.25em}-\lambda\chi(x)\hspace{-0.25em}+\hspace{-0.25em}\upsilon\big)\hspace{-0.25em}-\hspace{-0.25em}\mu(x)\hspace{-0.25em}-\hspace{-0.25em}\lambda_d\,J(x)^\top\hspace{-0.25em}\chi(x)\Big]
\end{align}
renders the set $\SM$ forward invariant and the closed-loop system $\delta$-ISS on $\SM$ with respect to $\upsilon$, with probability at least
$(1-\epsilon)^n$, if the gains $\lambda,\lambda_d>0$ satisfy
\begin{align}
    &\frac{\bar\upsilon^2}{\lambda^2}+\frac{\bar e^2}{2\lambda_d\lambda}\le M^2,\label{eqn:C1}\\
    &\lambda>L_D:=L_JL_eL_\psi+H_J\bar e+\lambda_dL_N,\label{eqn:C2}
\end{align}
where $\bar e$ and $L_e$ are given by~\eqref{eqn:gp_bound}
and~\eqref{eqn:Le}, $N(x):=J(x)J(x)^\top\chi(x)$, $L_\psi:=\max_{x\in\SM}\lVert J(x)^{-1}\rVert$, $L_J:=\max_{x\in\SM}\lVert J(x)\rVert$, $H_J:=\max_{z\in B_M}\Big\lVert\frac{\partial(J\circ\psi)}{\partial z}(z)\Big\rVert$, and $L_N:=\max_{z\in B_M}\Big\lVert\frac{\partial(N\circ\psi)}{\partial z}(z)\Big\rVert$.
\end{theorem}

\begin{proof}
We first prove that $\SM$ is forward invariant under the controller.
The control law~\eqref{eqn:ctrl} is well defined on $S$, since $J(x)$ is
invertible (Lemma~\ref{lem:chi}\ref{item: lemma_chi_i}). Substituting~\eqref{eqn:ctrl}
into~\eqref{eqn:sys_affine}, and using $g(x)g(x)^\dagger=\mathbb{I}$
(Assumption~\ref{ass:system}) we get,
\begin{align*}
    f(x)+g(x)u=\;&e(x)+J(x)^{-1}\big(\hspace{-0.25em}-\hspace{-0.25em}\lambda\chi(x)\hspace{-0.25em}+\hspace{-0.25em}\upsilon\big)\hspace{-0.25em}-\hspace{-0.25em}\lambda_dJ(x)^\top\chi(x).
\end{align*}
Since $\dot\chi=J(x)\dot x$ we have,
\begin{align}
    \dot\chi&=-\lambda\chi+\upsilon+\Xi(x),\label{eqn:chi_cl}
\end{align}
where $\Xi(x):=J(x)e(x)-\lambda_dN(x)$.

To analyse~\eqref{eqn:chi_cl}, we write it in terms of the transformed
coordinate $z=\chi(x)$. By Lemma~\ref{lem:chi}, $x=\psi(z)$, so
that~\eqref{eqn:chi_cl} becomes $\dot z=-\lambda z+\upsilon+\Xi(\psi(z))$.
Since $\psi$ maps $\R^n$ onto $S$, these dynamics are defined on all of
$\R^n$, and the right-hand side of $\dot z$ equation is locally Lipschitz in $z$. Hence,
for every $z(0)\in\R^n$, there is a unique solution on a maximal
interval $[0,T)$, $T\le\infty$, and the corresponding state
$x(t)=\psi(z(t))$ lies in $S$ for all $t\in[0,T)$. To show that $z(t)$
remains bounded, we consider the Lyapunov function
$V_s(z):=\frac12\lVert z\rVert^2$. Differentiating it, with $x=\psi(z)$,
we have
\begin{align*}
    \dot V_s&=z^\top\dot z
    =z^\top\big(-\lambda z+\upsilon+J(x)e(x)-\lambda_dJ(x)J(x)^\top z\big)\\
    &=-\lambda\lVert z\rVert^2+z^\top\upsilon+q^\top e(x)-\lambda_d\lVert q\rVert^2,
\end{align*}
where $q:=J(x)^\top z$. We bound the two indefinite terms separately.
Since $x\in S$, by Proposition~\ref{prop:gp} we have
$\lVert e(x)\rVert\le\bar e$ with probability at least $(1-\epsilon)^n$,
so by the Cauchy-Schwarz inequality and Young's inequality with the
gains $\lambda,\lambda_d$, we have
\begin{align*}
    \dot V_s&\le-\lambda \lVert z\rVert^2+\lambda_d\lVert q\rVert^2+\frac{\bar e^2}{4\lambda_d}+\frac{\lambda}{2}\lVert z\rVert^2+\frac{\bar\upsilon^2}{2\lambda}-\lambda_d\lVert q\rVert^2\\
    &=-\frac{\lambda}{2}\lVert z\rVert^2+\vartheta=-\lambda V_s+\vartheta,
\end{align*}
where $\vartheta:=\frac{\bar\upsilon^2}{2\lambda}+\frac{\bar e^2}{4\lambda_d}$.
By the comparison lemma, for all $t\in[0,T)$,
\begin{align*}
    V_s(t)\le e^{-\lambda t}V_s(0)+\frac{\vartheta}{\lambda}\big(1-e^{-\lambda t}\big)
    \le\max\Big(V_s(0),\frac{\vartheta}{\lambda}\Big).
\end{align*}
Since $V_s=\frac12\lVert z\rVert^2$ and, by~\eqref{eqn:C1},
$2\vartheta/\lambda=\bar\upsilon^2/\lambda^2+\bar e^2/(2\lambda_d\lambda)\le M^2$,
this gives
\begin{align}\label{eqn:chi_bound}
    \lVert z(t)\rVert\le\max\big(\lVert z(0)\rVert,M\big),\qquad\forall t\in[0,T).
\end{align}
Hence $z(t)$ remains in a bounded set and cannot escape in finite time,
so $T=\infty$~\cite[Theorem~3.3]{Khalil:1173048}. Moreover, if
$x(0)\in\SM$, i.e., $\lVert z(0)\rVert\le M$, then~\eqref{eqn:chi_bound}
gives $\lVert z(t)\rVert\le M$, i.e., $x(t)\in\SM$, for all $t\ge0$, so
$\SM$ is forward invariant.

Before proving $\delta$-ISS, we show that $\Xi$ is Lipschitz with
respect to the transformed coordinates on $\SM$. Let $x,x'\in\SM$,
$z:=\chi(x)$, $z':=\chi(x')$, and $\Delta\chi:=z-z'$, so that
$z,z'\in B_M$. For any continuously differentiable map $\Phi$ defined on
$S$, we have $\Phi(x)=(\Phi\circ\psi)(z)$, and since $B_M$ is convex,
Lemma~\ref{lem:mvt} applied to $\Phi\circ\psi$ on $B_M$ gives
\begin{align}\label{eqn:mvt_z}
    \lVert\Phi(x)-\Phi(x')\rVert\le\max_{\zeta\in B_M}\Big\lVert\frac{\partial(\Phi\circ\psi)}{\partial z}(\zeta)\Big\rVert\,\lVert\Delta\chi\rVert.
\end{align}
By the definition of $\Xi$, adding and subtracting $J(x)e(x')$,
\begin{align*}
    \Xi(x)-\Xi(x')=\;&J(x)\big(e(x)-e(x')\big)+\big(J(x)-J(x')\big)e(x')\\
    &-\lambda_d\big(N(x)-N(x')\big).
\end{align*}
We bound the three terms using~\eqref{eqn:mvt_z}. With $\Phi$ the
identity map, Lemma~\ref{lem:chi}\ref{item: lemma_chi_iii} gives
$\lVert x-x'\rVert\le L_\psi\lVert\Delta\chi\rVert$, and hence, by
Lemma~\ref{lem:lipschitz},
$\lVert J(x)(e(x)-e(x'))\rVert\le L_JL_e\lVert x-x'\rVert\le L_JL_eL_\psi\lVert\Delta\chi\rVert$.
With $\Phi=J$ and $\lVert e(x')\rVert\le\bar e$,
$\lVert(J(x)-J(x'))e(x')\rVert\le H_J\bar e\lVert\Delta\chi\rVert$.
With $\Phi=N$, $\lVert N(x)-N(x')\rVert\le L_N\lVert\Delta\chi\rVert$.
Combining these bounds,
\begin{align}\label{eqn:Xi_lipsc}
    \lVert\Xi(x)-\Xi(x')\rVert\le L_D\lVert\Delta\chi\rVert,
\end{align}
with $L_D$ as in~\eqref{eqn:C2}.

We now prove that the controller achieves $\delta$-ISS on $\SM$. Since
$\SM$ is compact and forward invariant, by Lemma~\ref{thm:lyapunov_char}
it suffices to show that the conditions of
Definition~\ref{def:deltaISSLyapunov}, with $A$ replaced by $\SM$, hold
for the $\delta$-ISS Lyapunov candidate
\begin{align*}
    V(x,x'):=\tfrac12\mathbf{d}(x,x')^2=\tfrac12\lVert\chi(x)-\chi(x')\rVert^2.
\end{align*}
Condition~\ref{item:lyapunov1} of Definition~\ref{def:deltaISSLyapunov} holds trivially with
$\underline\alpha(r)=\overline\alpha(r)=r^2/2$. To verify
condition~\ref{item:lyapunov2} of Definition~\ref{def:deltaISSLyapunov}, let $a,a'\in\SM$ be initial conditions,
$\upsilon,\upsilon':\R^+_0\rightarrow W$ be two external input signals,
$x(t):=x_{a\upsilon}(t)$, and $x'(t):=x_{a'\upsilon'}(t)$, both of which
remain in $\SM$. From~\eqref{eqn:chi_cl},
$\dot{\Delta\chi}=-\lambda\Delta\chi+\Delta\upsilon+\Xi(x)-\Xi(x')$ with
$\Delta\upsilon:=\upsilon-\upsilon'$. Using~\eqref{eqn:Xi_lipsc} we write,
\begin{align*}
    \dot V&=-\lambda\lVert\Delta\chi\rVert^2+\Delta\chi^\top\Delta\upsilon+\Delta\chi^\top\big(\Xi(x)-\Xi(x')\big)\\
    &\le-\kappa\lVert\Delta\chi\rVert^2+\Delta\chi^\top\Delta\upsilon,
\end{align*}
where $\kappa=\lambda-L_D>0$ by~\eqref{eqn:C2}. By Young's inequality,
$\Delta\chi^\top\Delta\upsilon\le\frac{\kappa}{2}\lVert\Delta\chi\rVert^2+\frac{1}{2\kappa}\lVert\Delta\upsilon\rVert^2$,
and therefore, using $\lVert\Delta\chi\rVert^2=2V$,
$\dot V\le-\kappa V+\frac{1}{2\kappa}\lVert\Delta\upsilon\rVert^2$, which is
condition~\ref{item:lyapunov2} with $\sigma(r):=r^2/(2\kappa)$. Since both conditions~\ref{item:lyapunov1}-\ref{item:lyapunov2} of
Definition~\ref{def:deltaISSLyapunov} are satisfied, the closed-loop
system is $\delta$-ISS on $\SM$ by Lemma~\ref{thm:lyapunov_char}.
Moreover, applying the comparison lemma to $\dot V$ and using
$\sqrt{p+q}\le\sqrt p+\sqrt q$, \eqref{eqn:deltaISS} holds with
$\beta(\mathbf{d}(a,a'),t):=\mathbf{d}(a,a')e^{-\kappa t/2}$ and
$\gamma(\lVert\upsilon-\upsilon'\rVert_\infty):=\frac{1}{\kappa}\lVert\upsilon-\upsilon'\rVert_\infty$.
Since the model error is bounded probabilistically, the closed-loop
system is $\delta$-ISS on $\SM$ and $\SM$ is forward invariant with
probability at least $(1-\epsilon)^n$.
\end{proof}

Unlike~\cite{sangeerth2025controller}, where the model errors at the two
trajectories are bounded separately and leave a constant offset in the
incremental bound, here the difference $e(x)-e(x')$ is bounded by
$L_e\lVert x-x'\rVert$, which vanishes as the trajectories approach each
other; hence the result is $\delta$-ISS rather than $\delta$-ISpS. If
$f$ is known, one can take $\mu=f$ and $\lambda_d=0$, so that
$L_D=0$, condition~\eqref{eqn:C1} reduces to $\bar\upsilon\le\lambda M$, and the result holds for every $M>0$. For
$\upsilon\equiv0$, the closed-loop equilibrium is then $x_c$; with a learned
model, the contraction of $\Delta\chi$ implies a unique closed-loop
equilibrium in $\SM$, with $\|\chi\|\le\bar e/(\kappa h(x_c))$. In either
case, $S$ need not contain an equilibrium of the open-loop system. The constants $L_\psi$,
$L_J$, $H_J$, and $L_N$ depend only on $h$, $x_c$, and $M$, and can be
computed offline by sampling $B_M$; the only quantity that depends on
the unknown $f$ is the RKHS bound $B_i$, as in all GP-based guarantees.
\begin{figure*}[t]
\centering
\includegraphics[width=\textwidth]{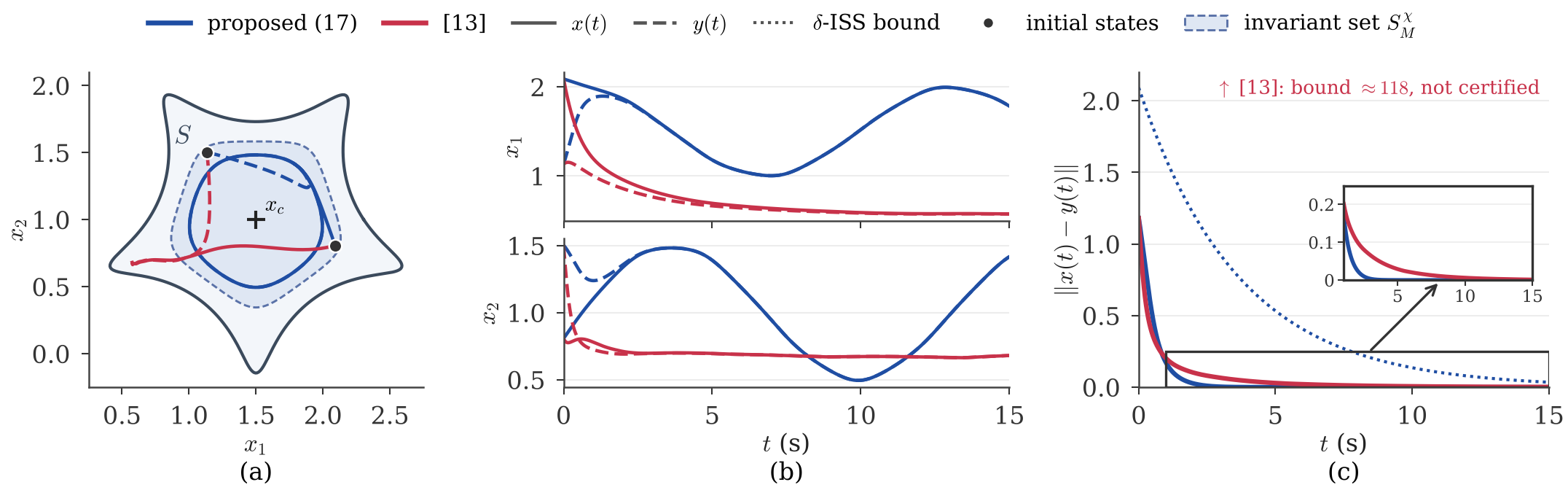}
\caption{Van der Pol oscillator: proposed controller~\eqref{eqn:ctrl} versus
\cite{basu2026neural}, from the same initial states and with the same input.
(a)~Trajectories in $S$. (b)~States. (c)~Distance $\|x-y\|$ and the certified
bound of~\eqref{eqn:ctrl} (dotted); the inset enlarges the boxed region.}
\label{fig:comparison}
\end{figure*}
\section{Case Studies}\label{sec:cases}

We illustrate Theorem~\ref{thm:main} on two systems whose safe sets exclude
every equilibrium of the open-loop drift, (i) a Van der Pol oscillator with a
non-convex safe set and (ii) a Lotka--Volterra system. In both cases, $f$ is
treated as unknown and learned from $N=1500$ noisy samples in $\bar S$, and
$\bar e$ and $L_e$ are computed from Proposition~\ref{prop:gp} and
Lemma~\ref{lem:lipschitz} with $\epsilon=10^{-3}$, i.e., with confidence
at least $0.998$.\footnote{The kernel hyperparameters are fixed before the
training data are collected, and $B_i$ is estimated from the simulation
model. Repeating the learning step on $200$ independent data sets, the bounds
of Propositions~\ref{prop:gp} and~\ref{prop:gp_deriv} were never violated.}

\subsection{Van der Pol Oscillator}

Let $f(x)=[x_2,\ 0.5(1-x_1^2)x_2-x_1]^\top$ and
$g(x)=\bigl[\begin{smallmatrix}1&0.2\sin x_2\\0&2+\cos x_1\end{smallmatrix}\bigr]$.
The safe set is defined, with $w=[w_1,w_2]=x-x_c$, $x_c=[1.5,1]^\top$, and $a=0.88$, by
\begin{equation*}
h(x)=1-\tfrac{\|w\|^2}{a^2}
-0.55\,\tfrac{5w_1^4w_2-10w_1^2w_2^3+w_2^5}{a^5}
-0.28\,\tfrac{\|w\|^6}{a^6}.
\end{equation*}
The resulting set $S$ has five lobes and is not convex
(Fig.~\ref{fig:comparison}(a)), but it satisfies
Assumption~\ref{ass:safe_set}.\footnote{Along the ray
$x=x_c+as[\cos\theta,\sin\theta]^\top$, \eqref{eqn:phi} gives
$\phi=1+s^2(1+2.2cs^3+1.4s^4)$ with $c=\cos(5(\theta-\pi/2))$, and the
bracket is positive on $S$, so $\phi\ge1$.}
With $M=1.25$, $\lambda=2$, and $\lambda_d=6\times10^{-5}$,
conditions~\eqref{eqn:C1} and~\eqref{eqn:C2} hold, with $\kappa=0.55$. Two trajectories start in different lobes of $\SM$ and receive the same
external input $\upsilon=\lambda r(t)$, where $r$ is a circular reference.
Both remain in $\SM$ and follow the reference
(Fig.~\ref{fig:comparison}(a)--(b)), and the distance between them decays to
zero, below the certified bound (Fig.~\ref{fig:comparison}(c)).

\emph{Comparison with~\cite{basu2026neural}.} We also applied the
neural-network design of~\cite{basu2026neural} to this example, with the
network structure and parameters reported there. Training took about
$21$\,min, whereas the complete design of~\eqref{eqn:ctrl} took $19$\,s including GP training.
The learned controller kept both trajectories in $S$, but its stability
conditions could not be certified. They were violated on about one third of
the test samples. In simulation, its trajectories converge more slowly
(Fig.~\ref{fig:comparison}(c)) and settle near a fixed point regardless of
$\upsilon$ (Fig.~\ref{fig:comparison}(b)), whereas under~\eqref{eqn:ctrl}
they follow the commanded reference.

\subsection{Lotka--Volterra System}

Let $f(x)=[x_1(1-0.5x_2),\ x_2(-0.75+0.25x_1)]^\top$ and
$g(x)=\mathrm{diag}(x_1,x_2)$, and let the safe set be defined by the concave
function $h(x)=1-(\tfrac{x_1-5}{2})^4-(\tfrac{x_2-4}{1.5})^4$, with
$x_c=[5,4]^\top$. The set excludes both equilibria of $f$, $[0,0]^\top$ and
$[3,2]^\top$. With $M=2$, $\lambda=1.5$, and $\lambda_d=2\times10^{-5}$,
conditions~\eqref{eqn:C1} and~\eqref{eqn:C2} hold, with $\kappa=1.10$.

\begin{figure}[t]
\centering
\includegraphics[width=0.75\columnwidth]{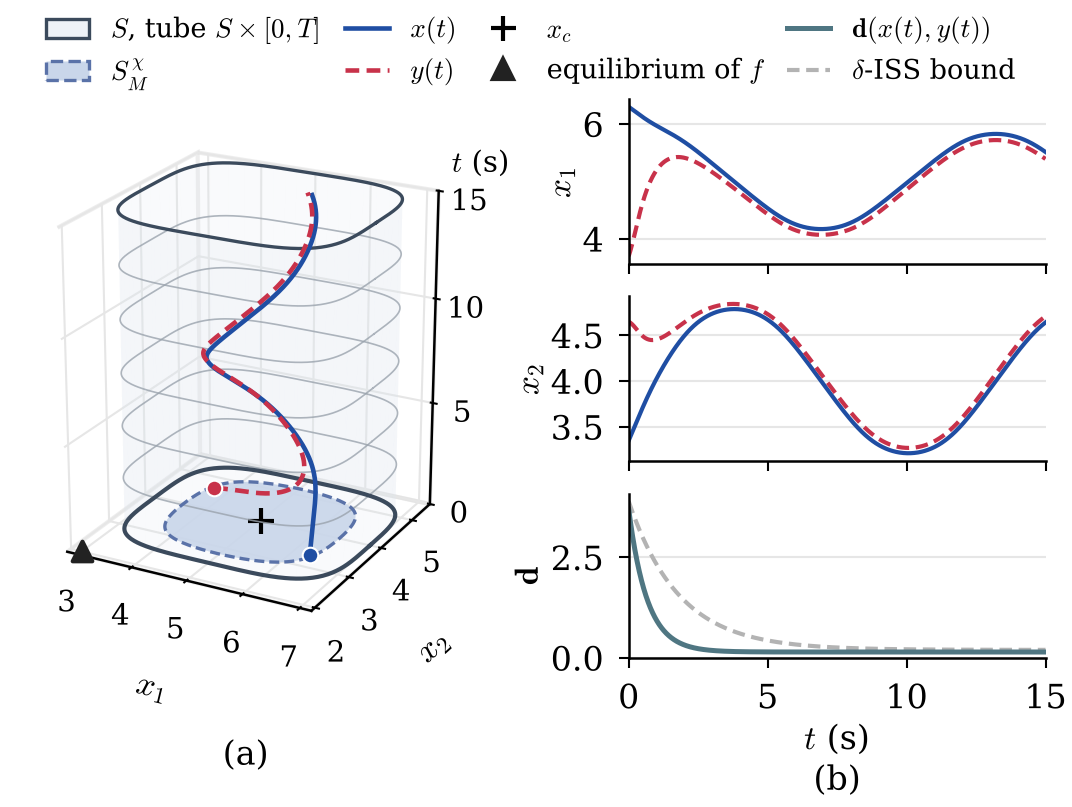}
\caption{Lotka--Volterra system. (a)~Trajectories in the set $S\times[0,T]$,
with $\SM$ at $t=0$. (b)~States, and distance $\mathbf d(x,y)$ with the
bound~\eqref{eqn:deltaISS}.}
\label{fig:lv}
\end{figure}

Here the two trajectories receive slightly different inputs, $\lambda r(t)$
and $\lambda(r(t)+\delta r)$, where $\delta r$ is a small constant offset.
Both remain in $\SM$, and their distance decays to a small value set by
$\delta r$, below the $\delta$-ISS bound (Fig.~\ref{fig:lv}). Unlike the
$\delta$-ISpS design of~\cite{sangeerth2025controller}, this residual distance is due only to the input difference and vanishes when the inputs
coincide.

Table~\ref{tab:comparison} compares the proposed design with existing
learning-based $\delta$-ISS controllers. Unlike~\cite{11312975}, it
guarantees safety; it is obtained in closed form, whereas the neural
controller of~\cite{basu2026neural} needs $20$\,min to $5$\,h of training
(comparative simulations shown in Fig.~\ref{fig:comparison}); and, compared
with~\cite{sangeerth2025controller}, it removes the practical offset and
keeps the state in the prescribed safe set rather than a relaxed one.
\begin{table}[t]
\centering
\caption{Comparison with existing learning-based $\delta$-ISS designs
(\ding{51}: yes, \ding{55}: no, --: not addressed).}
\label{tab:comparison}
\footnotesize
\setlength{\tabcolsep}{3pt}
\begin{tabular}{lcccccc}
\toprule
Approach & \shortstack{Safety} & \shortstack{No\\offset} & \shortstack{$x_{eq}\notin$\\safe set} & \shortstack{Noisy\\data} & \shortstack{Non-\\polyn.} & \shortstack{Low\\cost} \\
\midrule
\cite{11312975} (SOS, data-driven) & \ding{55} & \ding{51} & --        & \ding{55} & \ding{55} & \ding{51} \\
\cite{basu2026neural} (neural net.)  & \ding{51} & \ding{51} & --        & \ding{55} & \ding{51} & \ding{55} \\
\cite{sangeerth2025controller} (GP+CBF) & \ding{51} & \ding{55} & \ding{55}  & \ding{51} & \ding{51} & \ding{51} \\
Ours (GP+barrier map)              & \ding{51} & \ding{51} & \ding{51} & \ding{51} & \ding{51} & \ding{51} \\
\bottomrule
\end{tabular}
\end{table}

\section{Conclusion}\label{sec:conclusion}

We presented a feedback design for control-affine systems with an
unknown drift, learned via Gaussian process regression, that keeps the state in a prescribed safe set by rendering a compact subset of it forward invariant, and renders the closed-loop
system $\delta$-ISS with high probability, without the conservative practical offset of
previous learning-based designs. In particular, the safe set need not contain an equilibrium of the open-loop system.

\appendix

\subsection{Proof of Proposition~\ref{prop:gp_deriv}}
\label{appendix: proof_prop_gp_deriv}
\begin{proof}
Fix $i$. Let $\mathcal{H}$ be the RKHS of $k_i$ on $\hat S$,
$\varsigma=\rho_f^2$, $\Phi\xi=[\xi(x^{(1)}),\ldots,\xi(x^{(N)})]^\top$ for
$\xi\in\mathcal{H}$, $\Phi^*\boldsymbol{b}=\sum_{j}b_jk_i(\cdot,x^{(j)})$ for $\boldsymbol{b}\in\R^N$, and $\mathcal{V}=\Phi^*\Phi+\varsigma\mathrm{Id}$. The
bound of Proposition~\ref{prop:gp} is obtained
in~\cite[Theorem~3.11]{abbasi2012online} from the fact that, on the event $\mathcal{E}_i$ defined in the proof of Proposition~\ref{prop:gp}, $\lVert\mathcal{V}^{1/2}(f_i-\mu_i)\rVert_{\mathcal{H}}\le\sqrt{\varsigma}\,\eta_i$.
As $k_i$ is $C^2$ and $\hat S$ is open, $\ell=\partial k_i(\cdot,x)/\partial x_j\in\mathcal{H}$
and $\frac{\partial e_i}{\partial x_j}(x)=\langle\ell,f_i-\mu_i\rangle$~\cite{zhou2008derivative}.
By the Cauchy--Schwarz inequality,
$|\frac{\partial e_i}{\partial x_j}(x)|\le\lVert\mathcal{V}^{-1/2}\ell\rVert_{\mathcal{H}}\sqrt{\varsigma}\,\eta_i
=\eta_i\rho_{\partial,ij}(x)$, since
$\varsigma\lVert\mathcal{V}^{-1/2}\ell\rVert_{\mathcal{H}}^2=\rho_{\partial,ij}^2(x)$
by the matrix inversion lemma and~\eqref{eqn:deriv_var}. As
$\mathcal{E}_i$ does not depend on $x$ or $j$, this holds on $\mathcal{E}$.
\end{proof}

\subsection{Proof of Lemma~\ref{lem:chi}}
\label{appendix: proof_lem_chi}
\begin{proof}
\textit{(i)} Differentiating $\chi(x)=w/h(x)$ gives~\eqref{eqn:J}, which
has the form $J=A+ab^\top$ with $A=h^{-1}\mathbb{I}$, $a=-w/h^2$, and
$b=\nabla h$. Since $A^{-1}=h\mathbb{I}$, $A^{-1}a=-w/h$, and
$b^\top A^{-1}=h\nabla h^\top$, we have $A^{-1}ab^\top A^{-1}=-w\nabla h^\top$
and $1+b^\top A^{-1}a=1-\nabla h^\top w/h=\phi/h>0$. The
Sherman-Morrison formula then gives $J^{-1}=A^{-1}-\frac{A^{-1}ab^\top A^{-1}}{1+b^\top A^{-1}a}=h\,\mathbb{I}+\frac{w\nabla h^\top}{\phi/h}
    =h\Big(\mathbb{I}+\frac{w\,\nabla h^\top}{\phi}\Big).$

\textit{(ii)} Clearly $\chi(x_c)=0$. Every $x\in S\setminus\{x_c\}$ can be
written as $x=x_c+t\nu$ with $\lVert\nu\rVert=1$ and $t>0$, and then
$\chi(x)=F_\nu(t)\,\nu$ with $F_\nu(t):=t/h(x_c+t\nu)>0$. Thus $\chi$ maps
each ray from $x_c$ into the same ray, and it suffices to show that, on
each ray, $F_\nu$ maps the part of the ray inside $S$ one-to-one onto
$(0,\infty)$. Fix $\nu$. Since $S$ is bounded, there is a first $t^*>0$ with
$h(x_c+t^*\nu)=0$. Wherever the ray lies in $S$,
\begin{align*}
    F_\nu'(t)=\frac{h-t\,\nabla h^\top\nu}{h^2}=\frac{\phi(x_c+t\nu)}{h^2}>0,
\end{align*}
so $F_\nu$ is strictly increasing on every interval of the ray in $S$.
If the ray re-entered $S$ at some $t_2\ge t^*$, then $F_\nu(t)\to\infty$
as $t\downarrow t_2$, since $h\to0^+$, which contradicts this monotonicity.
Hence the ray lies in $S$ exactly for $t\in[0,t^*)$, so $S$ is star-shaped
with respect to $x_c$. Moreover, $F_\nu$ is continuous on
$[0,t^*)$, being a ratio of continuous functions with positive
denominator, with $F_\nu(0)=0$ and $F_\nu(t)\rightarrow\infty$ as
$t\rightarrow t^*$, since the numerator tends to $t^*>0$ and the
denominator to $0^+$. By the intermediate value theorem, $F_\nu$ attains
every value in $(0,\infty)$, and since it is strictly increasing, it
attains each value exactly once. Hence $\chi$ is a bijection from $S$
onto $\R^n$.

\textit{(iii)} Since $h\in C^2$ and $h>0$ on $S$, $\chi$ is $C^2$ on $S$.
By (i), $J(x)$ is invertible at every $x\in S$, so by the inverse function
theorem $\chi$ has a $C^1$ local inverse around every point, which by (ii)
coincides with $\psi$. Differentiating $\chi(\psi(z))=z$ gives
$J(\psi(z))\frac{\partial\psi}{\partial z}(z)=\mathbb{I}$.
\end{proof}

\bibliographystyle{ieeetr} 
\bibliography{sources} 

\end{document}